\documentclass[9pt,conference]{IEEEtran}

\usepackage{amsmath,amssymb,amsfonts}
\usepackage{algorithmic}
\usepackage[ruled,vlined,linesnumbered]{algorithm2e}
\usepackage{graphicx}
\usepackage{subcaption}
\usepackage{textcomp}
\usepackage{xcolor}
\usepackage[hidelinks]{hyperref}
\usepackage{url}
\usepackage{multirow}
\usepackage{xcolor}
\usepackage{listings}
\usepackage{comment}
\usepackage{xspace}
\usepackage{multirow}
\usepackage{colortbl}
\usepackage{combelow}
\usepackage{makecell}
\usepackage[utf8x]{inputenc}
\usepackage{balance}

\definecolor{mGreen}{rgb}{0,0.6,0}
\definecolor{myViolet}{rgb}{0.93,0.5,0.93}
\definecolor{mGray}{rgb}{0.5,0.5,0.5}
\definecolor{mPurple}{rgb}{0.58,0,0.82}
\definecolor{backgroundColour}{rgb}{0.95,0.95,0.92}
\definecolor{mGrayTbl}{RGB}{224, 224, 224}

\newcommand{\positEight}{\textit{Posit(8,1)}\xspace}
\newcommand{\positSixteen}{\textit{Posit(16,2)}\xspace}
\newcommand{\positThirtyTwo}{\textit{Posit(32,3)}\xspace}
\newcommand{\fpThirtyTwo}{\textit{FP32}\xspace}
\newcommand{\pau}{\textit{POSAR}\xspace}

\newcommand{\myor}{$ or $}
\newcommand{\myand}{$ and $}

\newcommand{\tablevspace}{\vspace{0pt}}

\lstdefinestyle{customc}{
belowcaptionskip=1\baselineskip,
breaklines=true,
frame=L,
xleftmargin=\parindent,
language=C,
showstringspaces=false,
basicstyle=\footnotesize\ttfamily,
keywordstyle=\bfseries\color{green!40!black},
commentstyle=\itshape\color{purple!40!black},
identifierstyle=\color{blue},
stringstyle=\color{orange},
morekeywords={uint32_t},
}

\def\BibTeX{{\rm B\kern-.05em{\sc i\kern-.025em b}\kern-.08em
T\kern-.1667em\lower.7ex\hbox{E}\kern-.125emX}}

\begin{document}

\pagenumbering{arabic}

\title{The Accuracy and Efficiency of Posit Arithmetic}

\author{\IEEEauthorblockN{Stefan Dan Ciocirlan$^{*+}$, Dumitrel Loghin$^*$,
Lavanya Ramapantulu, Nicolae \cb{T}\u{a}pu\cb{s}$^+$, Yong Meng Teo$^*$}\\
\IEEEauthorblockA{\textit{$^*$Department of Computer Science, National
University of Singapore, Singapore}\\
\textit{$^+$Department of Computer Science, University Politehnica
of Bucharest, Romania}\\
Email: $^*$\{dumitrel, teoym\}@comp.nus.edu.sg, lavanya.r@gmail.com,
$^+$\{stefandan, ntapus\}@cs.pub.ro
}}



\maketitle

\begin{abstract}

Motivated by the increasing interest in the posit numeric format, in this paper
we evaluate the accuracy and efficiency of posit arithmetic in contrast to the
traditional IEEE 754 32-bit floating-point (FP32) arithmetic. We first design
and implement a Posit Arithmetic Unit (PAU), called \pau, with flexible
bit-sized arithmetic suitable for applications that can trade accuracy for
savings in chip area. Next, we analyze the accuracy and efficiency of \pau with
a series of benchmarks including mathematical computations, ML kernels, NAS
Parallel Benchmarks (NPB), and Cifar-10 CNN. This analysis is done on our
implementation of \pau integrated into a RISC-V Rocket Chip core in comparison
with the IEEE 754-based Floting Point Unit (FPU) of Rocket Chip. Our analysis
shows that \pau can outperform the FPU, but the results are not spectacular. For
NPB, 32-bit posit achieves better accuracy than FP32 and improves the execution
by up to 2\%. However, \pau with 32-bit posit needs 30\% more FPGA resources
compared to the FPU. For classic ML algorithms, we find that 8-bit posits are
not suitable to replace FP32 because they exhibit low accuracy leading to wrong
results. Instead, 16-bit posit offers the best option in terms of accuracy and
efficiency. For example, 16-bit posit achieves the same Top-1 accuracy as FP32
on a Cifar-10 CNN with a speedup of 18\%.

\end{abstract}


\maketitle

\section{Introduction}

With the tremendous interest in Machine Learning (ML) which sparked new use
cases and business opportunities, companies are deploying Artificial
Intelligence (AI) models in users' devices, at the edge of the Internet.
Typical architectures used by these edge devices include ARM CPU cores
integrated with a Digital Signal Processor (DSP) and a Graphics Processing Unit
(GPU) to increase the performance of AI applications. In general, GPUs are more
time-energy efficient compared to CPUs when running ML applications, both during
the training and inference phases~\cite{dadiannao14}. But a recent study by
Facebook~\cite{mlfb_19}, shows that only 20\% of the smartphone SoCs have
accelerators that are significantly more efficient than the CPU. In addition, it is
more tedious to program accelerators compared to CPUs, due to different
architecture, programming model and, sometimes, due to the immaturity of the
Software Development Kit (SDK). Hence, it is important to optimize CPU inference
performance on mobile edge devices.

ML computations take place in two different stages, namely (i) at model training
when an ML model is built using huge datasets, and (ii) at inference when a new
sample or small dataset is processed by a trained model. Traditionally, ML
training is done on powerful accelerators, such as GPUs or Tensor Flow Units
(TPUs)~\cite{tpu}. To achieve high accuracy, ML models increasingly use large
numbers of parameters and, at runtime, they induce a huge number of arithmetical
operations. For example, VGG-16~\cite{simonyan2014very} from 2015 has
approximately 138 million parameters and 15,300 million multiply-add
operations~\cite{howard2017mobilenets, simonyan2014very}, while some current
submissions for ImageNet classification have one billion parameters. This
level of complexity is inappropriate for real-time applications, especially when
the inference is running on edge devices.

To address the complexity of ML inference, researchers have explored both model
pruning and lower bit representations for the parameters. Recent
studies~\cite{jeff_fb, deep_nn_gustafson} show that new floating-point
representations are more energy-efficient compared to the IEEE 754
floating-point standard implemented by all modern hardware processing units,
including CPUs and GPUs. IEEE 754 hardware implementations use significant chip
area and power because they need to handle many corner cases and exceptions
described by the standard. Some studies show that that this standard is
error-prone~\cite{gustafson2015end} and its different implementations may produce
different results~\cite{Whitehead2011PrecisionAP}.

Among the alternatives to IEEE 754, \textit{unum}~\cite{gustafson2015end} and
its third version, named \textit{posit}~\cite{gustafson2017beating,posit_standard},
were introduced by Gustafson to
solve some of the issues of IEEE 754 floating-point representation. Compared to
IEEE 754, posit has variable length fields to represent the exponent and the
fraction of a real number. Hence, posits can represent small numbers more
accurately by reserving more bits for the fraction and fewer bits for the
exponent. Moreover, posits have only two special representations, namely for $0$
and not-a-real ($NaR$), whereas IEEE 754 reserves many binary representations
for not-a-number ($NaN$). This feature, together with the non-existence of
subnormal make posit implementations simpler.
In this paper, we address the following research questions, (i) are posits more
efficient than IEEE 754 for ML inferencing at the edge? (ii) what is a good
trade-off between accuracy and time-energy efficiency when employing lower bit
size posits?

To answer these questions, we propose an alternative hardware-software approach
for efficient ML inference at the edge by (i) designing and implementing an
Elastic Posit Arithmetic Unit (\pau) in a Rocket
Chip-based~\cite{asanovic2016rocket} RISC-V core replacing its IEEE 754
floating-point unit (FPU), and (ii) modifying existing software to run on this
system.
RISC-V~\cite{waterman2016design} is an open-source architecture with limited
available hardware implementations. However, it is very promising due to its
energy efficiency and modular ISA, thus, it is timely to explore this new CPU
architecture for ML inference at the edge. Without modifying the ISA, we use the
$F$ extension of the RISC-V specification~\cite{waterman2019isa} but change the
internal processor representation of floating-point numbers to posit. To address
the challenge of executing the same application software on the modified
hardware, we make minor high-level code changes and convert IEEE 754 constants
to posit.

To evaluate the \pau, we use three application levels and compare their
accuracy, execution time, and power consumed on an FPGA implementation. In
addition, we compare the resource utilization of \pau on the FPGA. Among the
three application levels, the first consists of computing the values of
well-known mathematical constants, such as $\pi$ and $e$ for testing accuracy
over a large number of operations. The second level represents kernels
frequently used in ML applications, such as matrix multiplication, k-means, and
linear regression, among others. The third level is a Convolutional Neural
Network (CNN) running on Cifar-10 dataset.

Among the existing posit hardware
implementations~\cite{chaurasiya2018parameterized, 8351142, 8342187, 8425396,
8731915}, most are accelerators, co-processors or stand-alone units. The closest
work to ours is PERI \cite{peri_riscv_core}, which designs a PAU for the RISC-V
core SHAKTI implemented in Bluespec System Verilog for 32-bit posits.
However, PERI is evaluated only on 32-bit posits with two exponent sizes while
running image processing, FFT, trigonometric functions, and K-means. In
contrast, our proposed approach is \textit{elastic} as it can be adapted to
different posit sizes. In this paper we have evaluated \pau on three posit
sizes of 8, 16, and 32 bits. To the best of our knowledge, we are the first to
(i) implement an \pau in the Chisel language (a dialect of Scala) and integrate
it into a Rocket Chip core, and (ii) evaluate a posit-enabled Rocket Chip system
on an FPGA. In addition, we release the code associated with our project and the
documentation as open-source to aid future research in this
area\footnote{\url{https://github.com/dloghin/posar}}.

In summary, we make the following contributions:
\begin{itemize}

\item We implement an Elastic Posit Arithmetic Unit (\pau) in the Chisel
language to replace the traditional Floating Point Unit (FPU) in a Rocket Chip
RISC-V core. In addition, we implement a Scala library for posit arithmetic to
test our \pau.

\item Our \pau supports any posit size and exponent size, but in this paper
we instantiate it for three sizes: 8, 16, and 32 bits, respectively.
We test these instantiations in both simulation mode and on an Arty A7-100T
FPGA.

\item  Our evaluation shows that \pau with 32-bit posit achieves the same
accuracy as the original FPU of Rocket Chip with 32-bit IEEE 754
floating-point. \pau is logging fewer cycles, but uses more FPGA resources and
consumes more power. However, our energy measurements on the FPGA running a
loop of two million iterations that computes $\pi$ show that 32-bit posit uses
only 6\% more energy while being 30\% faster compared to 32-bit IEEE 754
float.

\item We show that 8-bit posit is not suitable to replace 32-bit IEEE 754
float in scientific and classic ML applications due to low accuracy. On
the other hand, 16-bit posit offers the best option in terms of accuracy and
efficiency. For example, 16-bit posit achieves the same Top-1 accuracy as
32-bit IEEE 754 single-precision format on a Cifar-10 CNN with a speedup of
18\%. Moreover, a hybrid approach where 8-bit posit is used to store
parameters in memory and a 16-bit posit \pau is used for computations leads to
no loss of accuracy on the Cifar-10 CNN.

\end{itemize}

In the next section, we provide details on ML complexity, posit and RISC-V. In
Section~\ref{sec:relwork} we summarize related works. In
Section~\ref{sec:approach} we present the design of our \pau and discuss the
challenges of enabling existing applications to run on it. In
Section~\ref{sec:eval} we evaluate our approach before concluding in
Section~\ref{sec:concl}.

\section{Background}
\label{sec:background}

\subsection{Machine Learning Complexity}

As previously stated, ML applications are becoming increasingly complex from the
point of view of parameters and runtime operations. One way to address this
issue is pruning which aims to reduce the depth and complexity of a neural
network. Howard et al.~\cite{howard2017mobilenets} propose the usage of
depth-wise and point-wise convolutions instead of full convolution to reduce the
number of parameters by 32 times and the number of operation by 27 times in
comparison with the original VGG-16 while achieving nearly the same accuracy.
Another approach is knowledge distillation~\cite{mishra2017apprentice} where a
smaller network is trained with a larger network in a teacher-student style.
This method achieves better accuracy than the case where the small network is
trained alone. A third approach is quantization~\cite{quantization_16} which
replaces floating-point arithmetic with other numeric representations, such as
fixed-point, to reduce the number of bits and the cost of operations in terms of
latency.

The quantization approach uses different types of numerical representations,
such as binary, ternary, integer, fixed-point, small-size floating-point, and
posit~\cite{gustafson2017beating}. Depending on its usage, quantization can be
classified into storage or computing. The storage quantization tackles the
problem of memory usage by adopting a lower bit representation in memory for
integers, fixed-point~\cite{sakr2017analytical}, floating-point and posit
numbers~\cite{langroudi2018deep}. But before performing the operations, these
lower bit representations are converted to full-size floating-point.
This method degrades accuracy and increases latency.
To address this issue, Langroudi et al.~\cite{langroudi2018deep} proposed using
different bit sizes per ML network layer. This method decreases memory
utilization by as much as 94\% while degrading the accuracy by only 1\%. This is
a good solution for saving storage but it may increase design complexity.

Carmichael et al.~\cite{deep_nn_gustafson} analyze fixed-point,
floating-point and posit representations on DNN showing that posit offers the
best accuracy while exhibiting a smaller latency compared to floats. In this
paper, we aim to explore deeper the suitability of using posit for ML inference.
Specifically, we evaluate three different posit sizes to determine which one
offers the best accuracy-efficiency trade-off.

\subsection{Posit}
\label{sec:posit}

\begin{figure}[tp]
\centering
\includegraphics[width=0.475\textwidth]{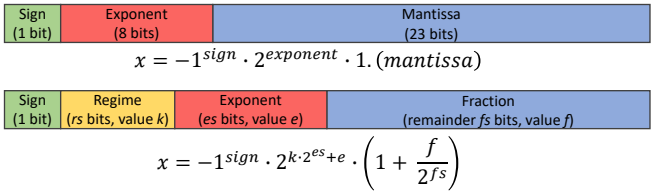}
\caption{IEEE 754 Single-precision floating-point (top) vs. posit (bottom)}
\label{fig:numberformats}
\end{figure}

Posit~\cite{posit_standard} is a real number representation that aims to improve
the widely-used IEEE 754 floating-point standard implemented by the majority of
modern processors. A 32-bit, single-precision floating-point in IEEE 754
comprises three fields, as shown in Figure~\ref{fig:numberformats} (top), namely
(i) a sign field of 1 bit, (ii) an exponent field of 8 bits and (iii) a fraction
or mantissa field of 23 bits. In contrast to IEEE 754 which reserves many binary
representations for the special number $NaN$, posit format has only two special
numbers, $0$ and $NaR$ (not-a-real). If the binary representation of the posit
has all the bits equal to zero, except the first bit from the left which
represents the sign, then it is a special number. If the sign bit is $0$ then
the special posit has the value $0$, otherwise it represents the posit $NaR$, as
shown in Table~\ref{tab:posit_ex} for 8-bit posit.
Compared to IEEE 754, posit representation comprises an additional field named
\textit{regime} which determines the final exponent value together with the
exponent field, as shown in Figure~\ref{fig:numberformats} (bottom). In posit,
the regime and fraction fields are variable, while the exponent field is
customizable. In fact, a posit format can be described only by its total size,
$ps$, and its exponent size, $es$.

\begin{equation}
\label{eq:k}
k = \left \{
\begin{aligned}
&-rn, && \text{if}\ r_{i}=0 \\
&rn-1, && \text{otherwise}
\end{aligned} \right.
\end{equation}

In a posit, the regime field follows the 1-bit sign and continues as long as the
bits have the same value $r_i$, followed by a bit of opposed value. The number
of regime bits of the same value, $rn$, is used to determine the value $k$ as
shown in Equation~\ref{eq:k}. $k$ is a factor that multiplies the maximum value
of the exponent field, $2^{es}$, to which the actual value of the exponent
field, $e$, is added to determine the final exponent ($k\cdot2^{es}+e$).
This feature of elastic exponent field allows a larger fraction field, hence, a
higher representation accuracy compared to the fixed 23-bit mantissa of the IEEE
754 format. However, this higher accuracy occurs only in a range called the
``golden zone'' \cite{de2019posits} which can be useful in scientific
applications.

\begin{table}[tp]
\centering
\caption{Examples of 8-bit Posits with 1-bit exponent. The colors represent the fields as in Figure~\ref{fig:numberformats}.}
\label{tab:posit_ex}
\begin{tabular}{|r|r|}
\hline
\multicolumn{1}{|c|}{\textbf{Value}} & \multicolumn{1}{c|}{\textbf{Binary Representation}} \\
\hline
\hline
0 & \texttt{\color{olive}0~\color{black}0 0 0 0 0 0 0} \\
NaR & \texttt{\color{olive}1~\color{black}0 0 0 0 0 0 0} \\
1.0 & \texttt{\color{olive}0~\color{orange}1 0~\color{red} 0~\color{blue}0 0 0 0} \\
-2.0 & \texttt{\color{olive}1~\color{orange}0 1~\color{red} 1~\color{blue}0 0 0 0} \\
3.125 & \texttt{\color{olive}0~\color{orange}1 0~\color{red} 1~\color{blue}1 0 0 1} \\
\hline
\end{tabular}
\end{table}

To improve the rounding error and to abide by mathematical properties such as
associativity and distributivity, the posit standard introduces a
\textit{quire}~\cite{posit_standard} which is a long
accumulator~\cite{de2019posits}. However, the implementation of a quire uses 10
times more area and increases the latency by 8 times compared to posit without
quire~\cite{de2019posits}. For example, an implementation of an unum type
co-processor in SMURF~\cite{bocco2019smurf} for a RISC-V Rocket-Chip uses 9
times more area and consumes 12 times more energy than the 64-bit FPU of the
Rocket-Chip. Thus, in this paper, we decided not to implement a quire in our
\pau.

\subsection{RISC-V and Elastic ISA}
\label{sec:riscv}

RISC-V is a Reduced Instruction Set Computer (RISC) instruction set architecture
(ISA) that is open-source and designed in a modular way that permits adding new
extensions~\cite{waterman2016design}. A RISC-V processor design must start with
a base ISA module, such as 32- or 64-bit integer operations to which other
modules can be added. For example, single- and double-precision floating-point
extensions, denoted by letters F and D, respectively, can be added to a
processor to support floating-point arithmetic. There is a multitude
of open-source implementations of RISC-V architecture, some of which are
documented on the GitHub page maintained by the RISC-V
community~\cite{riscv_git}.

Recently, even commercial architecture companies, like ARM, have opened up their
architectures for customized extensions by partners and have added custom
instructions to their ISA~\cite{arm-custom}. These developments in the industry
are in line with the research moving towards ``elastic ISA''.
Such an elastic ISA provides the core set of instructions for a domain-specific
application such that the processor's hardware units are efficiently utilized in
order to achieve minimal energy usage for a given execution time performance
target. This is directly applicable in the field of ML due to the fast pace of
development of algorithms versus the pace of available hardware catering to the
throughput and power demands of inference at the edge~\cite{zhou2019edge}. The
\pau proposed in this work helps bridge this gap between algorithms and
hardware as the elasticity in its compute engine lends itself easily to advances
in ML algorithms in the area of changing arithmetic representations.

\section{Related Works}
\label{sec:relwork}

\textbf{Posit Implementations in Hardware.} The closest work to ours is PERI, a
posit-enabled RISC-V core presented in a preprint~\cite{peri_riscv_core}.
Similar to our work, PERI implements a posit unit capable of executing RISC-V F
extension instructions, presents a similar way to run existing floating-point
programs on the posit-enabled processor, and evaluates the processor on an Arty
A7-100T FPGA. However, we implement our project in the Chisel language to
integrate it with the Rocket Chip core, while PERI uses Bluespec System Verilog
and integrates the unit into the SHAKTI core~\cite{SHAKTI_16}.
PERI uses two posit formats at the same time. Both are 32-bit in size, with one
having $es = 2$ and the other $es = 3$. To switch between these two formats,
PERI introduces a new instruction, $FCVT.ES$. In contrast, our \pau supports
multiple bit-sized posits but uses only one size at a time to keep full
compatibility with existing software. We evaluate 8- and 16-bit posits in
addition to 32-bit posits in this paper. In terms of benchmarking, both PERI and
we use k-means, along with the computation of $sin(x)$ and $e$. We evaluate $pi$
computation, matrix multiplication, k nearest neighbors, naive Bayes, and
support vector machine, while PERI is evaluated on JPEG image processing and
fast Fourier transform.

Other works propose incomplete posit arithmetic
units~\cite{chaurasiya2018parameterized, 8351142, 8342187, 8425396, 8731915}.
Specifically, \cite{chaurasiya2018parameterized} presents a hardware generator
that can produce posit adders and multipliers, \cite{8351142} presents an
adder/subtractor, \cite{8425396} presents an adder, subtractor, and multiplier,
while \cite{8342187} implements conversion of IEEE 754 to/from posit, adder,
subtractor, and multiplier. Compared to \cite{8342187}, \cite{8731915} also
implements a divider. Similar to us, \cite{chaurasiya2018parameterized} show
that posit unit operating on the same size (e.g. 32 bits) as an IEEE
754-compliant unit needs slightly more hardware resources. In contrast,
\cite{8425396} observes that posit takes significantly more FPGA resources than
IEEE 754.

\textbf{Posit in Machine Learning.} Some works analyze the suitability of using
posits in ML applications~\cite{deep_positron_arxiv, deep_nn_gustafson,
Langroudi18, jeff_fb}. \cite{deep_nn_gustafson} and \cite{deep_positron_arxiv}
present Deep Positron, an Deep Neural Network (DNN) accelerator that can run on
FPGAs, and claim that 8-bit posits can achieve better inference accuracy than
8-bit floats and integers, while being close to the accuracy of 32-bit IEEE 754
floats. In contrast, we find that 8-bit posits almost always produce inaccurate
results when compared to both 32-bit posits and IEEE 754 floats. However, we
acknowledge that we use Cifar-10 dataset for the ML application, whereas Deep
Positron uses medical low-dimensionality datasets. Johnson~\cite{jeff_fb}
evaluates multiple numeric formats, including posit, to replace IEEE 754 floats
in DNNs. Among others, the author shows that 8-bit posits with 1-bit exponents
can achieve similar Top-1 accuracy on a Resnet50 model compared to classic
32-bit floats. In \cite{Langroudi18}, posits are used to store ML parameters in
memory, being converted to classic IEEE 754 floats when computations are
performed. The authors claim that posits of smaller size can represent the
parameters compared to bigger sized IEEE 754 floats, and hence, save up to 36\%
of memory space while less than 1\% accuracy loss is exhibited. In contrast, we
show that frequent conversions between posit and IEEE 754 formats can lead to
significant accuracy degradation and discuss this in detail in
Section~\ref{sec:software}.

\textbf{Posit in Scientific Computing.} In \cite{posit_npb}, the authors
evaluate the impact of posit on NPB benchmarks using software emulation. As
expected, the emulation leads to a much higher execution time of the program
using posit compared to the IEEE 754 format running natively on the hardware.
However, the accuracy of 32-bit posit is higher compared to FP32. In
\cite{gustafson2017beating}, the authors show using high-level emulation that
32-bit posit can achieve better accuracy and (potentially) faster execution
compared to the IEEE 754 format on the LINPACK benchmark. In contrast, we use a
hardware-based approach to better understand the impact of posit on
cycle-efficiency, not only on accuracy.

\begin{figure*}
\centering
\includegraphics[width=0.6\textwidth]{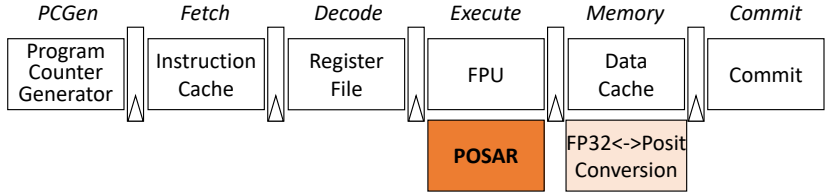}
\caption{\pau in a Rocket Core pipeline}
\label{fig:rocketchip_pipeline}
\end{figure*}

\section{\pau: Design and Implementation}
\label{sec:approach}

In this section, we describe the implementation of \pau and the high-level
design choices we made to run existing software on the system that integrates
\pau.

\subsection{\pau}

In this section, we describe the implementation of \pau, our Posit Arithmetic
Unit. We wrote the high-level design code in the Chisel language and integrated
it into a Rocket Chip tiny core to replace the original Floating Point Unit
(FPU) than implements the IEEE 754 standard. Our \pau is activated during the
execution phase of the pipeline, as shown in
Figure~\ref{fig:rocketchip_pipeline}. In addition to supporting all the
instructions of the F extension of RISC-V~\cite{waterman2019isa}, our \pau is
elastic to cater to parameterized sizes for posit and exponent. Using this
elastic feature, we evaluate \pau on 8-, 16-, and 32-bit posits ($ps$) with 1-,
2-, and 3-bit exponents ($es$), respectively. To verify our implementation, we
wrote a posit arithmetic library in Scala and tested the \pau Chisel code using
unit testing. In this section, we discuss our internal posit representation, the
instructions implemented, and the challenges of running programs on a
posit-enabled processor.

\begin{table}[tp]
\centering
\caption{Notations}
\label{table:notations}
\resizebox{0.48\textwidth}{!} {
\begin{tabular}{|r|l|}
\hline
\textbf{Symbol} & \multicolumn{1}{c|}{\textbf{Description}} \\
\hline
\hline
$P$ & posit number \\
$BP$ & posit binary representation. The right-most bit index is 1. \\
$ps$ & posit size \\
$es$ & exponent size \\
$sn$ & special number bit, set for $0$  and $NaR$ \\
$s$ & sign \\
$rn$ & number of regime bits with the same value \\
$rs$ & regime size, $rs=rn+1$ \\
$r_{i}$ & regime bits with the same value \\
$\overline{r}$ & last regime bit with a different value, $\overline{r}=1-r_{i}$
\\
$k$ & regime value \\
$e$ & exponent value \\
$ers$ & exponent size in $BP$, $ers = max(0, min(es, ps - rs - 1))$ \\
$fs$ & fraction size \\
$f_{i}$ & fraction bits \\
$f$ & fraction value \\
$frs$ & fraction size in $BP$, $frs = max(0,ps-rs-es-1)$ \\
\hline
\end{tabular}
}
\tablevspace
\end{table}

\textbf{Supported Instructions.} \pau supports all the instructions of the F
extension of RISC-V~\cite{waterman2019isa}. For bitwise addition, subtraction,
multiplication, and division we used the Chisel build-in operators. We
acknowledge that this choice leaves some room for further optimizations.
For testing the hardware implementation, we wrote a small library in Scala
representing posit numbers and we used it inside unit tests. We hope this
library will help others in trying different hardware implementations of posit
operations.

\begin{algorithm}[tp]
\caption{Posit Decoding}
\label{alg:posit_decode}
\SetKwFunction{LeadingOnes}{LeadingOnes}
\SetKwFunction{Reverse}{Reverse}
\SetKwFunction{Cat}{Cat}
\SetKwInOut{Input}{input}\SetKwInOut{Output}{output}
\Input{$BP, ps, es$}
\Output{$s,sn,k,rs,e,ers,f,fs$}
\BlankLine

$sn \leftarrow \sim |BP[ps-1:1]$\;~\tcc{\small$\sim |$ is OR-NOT operation on
all bits}

$s \leftarrow BP[ps]$ \;
\lIf{$s=1$}
{$BP \leftarrow \sim BP + 1$}

$r_{i} \leftarrow BP[ps-1]$;

\If{$r_{i}=1$}
{
$rn \leftarrow$ \LeadingOnes{\Reverse{$BP$}}\;
$k \leftarrow rn-1$;
}
\Else
{
$rn \leftarrow$ \LeadingOnes{\Reverse{$\sim BP$}}\;
$k \leftarrow -rn$;
}
$rs \leftarrow rn+1$\;

$ers \leftarrow max(0,min(es,ps-rs-1))$\;
\lIf
{$ers=0$}
{$e \leftarrow 0$}
\lElse
{$e \leftarrow BP[(ps-1-rs):(ps-rs-ers)] << (es-ers)$}

$frs \leftarrow max(0,ps-rs-es-1)$\;
\lIf{$frs=0$}{$f \leftarrow 0$}
\lElse{$f \leftarrow BP[(ps-1-rs-es):1]$}
$f \leftarrow f + 2^{fs}$
\end{algorithm}

\begin{algorithm}[tp]
\caption{Posit Encoding}
\label{alg:posit_encode}
\SetKwFunction{Reverse}{Reverse}
\SetKwFunction{Cat}{Cat}
\SetKwInOut{Input}{input}\SetKwInOut{Output}{output}
\Input{$ps, es, s, sn, k, e, f, fs, bm$}
\Output{$BP$}
\BlankLine
\If{$sn=1$}
{
\lIf{$s=1$}
{$BP \leftarrow 1 \ll (ps-1)$}
\lElse
{$BP \leftarrow 0$}
}
\Else
{
\If{$k \geq (ps-2)$}
{
\tcc{\small{if value is bigger than maxvalue then $BP \leftarrow
maxvalue$}}
$BP \leftarrow (1 \ll (ps-1))-1$

}
\ElseIf{$k< -(ps-2)$}
{
\tcc{\small{if value is smaller than minvalue then $BP \leftarrow
minvalue$}}
$BP \leftarrow 1$
}
\Else
{
\If{$k \geq 0$}
{
$rn \leftarrow k+1$\;
$regimebits \leftarrow ((1 \ll rn)-1) \ll 1$\;
$rs \leftarrow rn+1$\;
}
\Else
{
$rn \leftarrow-k$\;
$regimebits \leftarrow 1$\;
$rs \leftarrow rn+1$\;
}

$nrs \leftarrow max(0,ps-rs-1)$\;

$regimebits \leftarrow regimebits \ll
nrs$\;

$f \leftarrow f \ll (2 \cdot ps-fs)$\;

$othervalue \leftarrow$ \Cat{$e$, $f[2 \cdot ps:1]$} $ \ll (ps-es)$ \;

$otherbits \leftarrow othervalue[3 \cdot ps:2 \cdot ps+1] \gg (ps-nrs)$\;

$BP \leftarrow (regimebits | otherbits)$\;

$b_{n+1} \leftarrow (othervalue[3 \cdot ps:2 \cdot ps+1] \gg (ps-nrs-1))
\& 1$\;

$bm \leftarrow (|(othervalue[3 \cdot ps:2 \cdot ps+1] \& ((1 \ll (ps-nrs-1))-1))) | (|(othervalue[2
\cdot ps:1]))  | bm$\;

$addOne \leftarrow b_{n+1} \& (bm | (\sim bm \& BP[1]))$\;

$BP \leftarrow BP + addOne$\;
}
\lIf{$s=1$}{$BP \leftarrow \sim BP + 1$}
}
\end{algorithm}

\textbf{Posit Representation.} We use an internal posit representation
comprising the sign $s$, the regime $k$ and its size $rs$, the exponent $e$ and
its actual size in the binary representation, $ers$, the fraction $f$ and its
size $fs$, and one bit $sn$ for the special numbers $0$ and $NaR$. We decode a
binary posit representation before performing an operation in the \pau using
Algorithm~\ref{alg:posit_decode}. We encode our internal representation into a
binary representation at the end of the operation, using
Algorithm~\ref{alg:posit_encode}. The notations used by the algorithms presented
in this section are summarized in Table~\ref{table:notations}.

In addition to making it easy to implement the arithmetic operations, this
internal posit representation allows us to keep the additional bit resulted
after applying different operations on the posit. In turn, this bit helps us
perform better rounding when encoding the binary posit representation.

\textbf{Decoder and Encoder.} The posit decoder takes as input a binary
representation, $BP$, the posit and exponent sizes, $ps$ and $es$, respectively,
and computes $s,sn,k,rs,e,ers,f,fs$, as shown in
Algorithm~\ref{alg:posit_decode}.
The algorithm verifies if $BP$ represents one of the special numbers,
$0$ or $NaR$.
In this case, we set $sn$ and $s$. If $BP$ is not a special number and $s$ is
set, it represents a negative number and we take its two's complement.
Based on the first bit of the regime, $r_{i}$, we compute $rn$ and $k$. If
$r_{i}$ is $0$, all bits of the $BP$ are negated, then reversed, and the result
goes to a leading ones detector which computes $rn$. $k$ is then computed based
on Equation~\ref{eq:k}. If $r_{i}$ is $1$, the same steps are applied, except
that $BP$ is not negated.

\begin{algorithm}[tp]
\caption{Posit Add/Sub Selector}
\label{alg:posit_adder_selector}
\SetKwInOut{Input}{input}\SetKwInOut{Output}{output}
\Input{$P_{1}, P_{2}, op$}
\Output{$P_{1}, P_{2}, op, sign$}
\BlankLine
\If{$op=0$}
{
\lIf{$P_{1}.s=P_{2}.s$}
{
$sign \leftarrow P_{1}.s$
}
\Else
{
\If{$P_{1}.s=1$}
{
$op \leftarrow 1$\;
$sign \leftarrow 1$\;
}
\Else
{
$op \leftarrow 1$\;
$sign \leftarrow 0$\;
}
}
}
\Else
{
\lIf{$P_{1}.s=P_{2}.s$}
{
$sign \leftarrow P_{1}.s$
}
\Else
{
\If{$P_{1}.s=1$}
{
$op \leftarrow 0$\;
$sign \leftarrow 1$\;
}
\Else
{
$op \leftarrow 0$\;
$sign \leftarrow 0$\;
}
}
}
\If{$abs(P_{1})<abs(P_{2})$}
{
$aux \leftarrow P_{1}$\;
$P_{1} \leftarrow P_{2}$\;
$P_{2} \leftarrow aux$\;
\lIf{$op=1$}
{
$sign \leftarrow 1 - sign$
}
}
\end{algorithm}

\begin{algorithm}[tp]
\caption{Posit Adder/Subtractor}
\label{alg:posit_adder}
\SetKwFunction{PositAddSubSelector}{PositAddSubSelector}
\SetKwInOut{Input}{input}\SetKwInOut{Output}{output}
\Input{$P_{1}, P_{2}, op$}
\Output{$P_{3}$}
\BlankLine

$P_{1},P_{2},op, P_{3}.s \leftarrow~$\PositAddSubSelector~($P_{1},P_{2},op$);

\lIf {$(P_{1}.sn=1 \myand P_{1}.s=1) \myor (P_{2}.sn=1 \myand P_{2}.s=1)$}
{
$P_{3} \leftarrow NaR$
}
\lElseIf{$P_{2}.sn=1 \myand P_{2}.s=0$}
{
$P_{3} \leftarrow P_{1}$
}
\Else
{
$P_{3}.s \leftarrow P_{1}.s$\;
$P_{3}.sn \leftarrow 0$\;
$P_{3}.k \leftarrow P_{1}.k$\;
$P_{3}.e \leftarrow P_{1}.e$\;
$P_{3}.es \leftarrow P_{1}.es$\;
$P_{3}.fs \leftarrow (2 \cdot ps-4)$\;
$t \leftarrow (P_{1}.k \ll es + P_{1}.e) - (P_{2}.k \ll es + P_{2}.e)$\;
\lIf{$op=0$}
{
\tcc{\small{add}}
$P_{3}.f  \leftarrow (P_{1}.f \ll (P_{3}.fs-P_{1}.fs)) + ((P_{2}.f \ll (P_{3}.fs-P_{2}.fs)) \gg t)$
}
\lElse
{
\tcc{\small{subtract}}
$P_{3}.f  \leftarrow (P_{1}.f \ll (P_{3}.fs-P_{1}.fs)) - ((P_{2}.f \ll (P_{3}.fs-P_{2}.fs)) \gg t)$
}
$P_{3}.bm = |((P_{2}.f \ll (P_{3}.fs-P_{2}.fs)) \& ((1 \ll t)-1))$\;
}
\end{algorithm}

The posit encoder takes as input $ps, es, s, sn, k, e, f, fs, bm$ and computes
$BP$, as shown in Algorithm~\ref{alg:posit_encode}. Here, $bm$ is a bit which is
set when bits of value one are present in the extended fraction, after the first
$fs$ bit. This bit affects the final value of $BP$, as we shall see below. If
$sn$ is set, $BP$ is one of the special numbers $0$ or $NaR$, depending on $s$.
Otherwise, Algorithm~\ref{alg:posit_encode} checks if the regime is greater or
equal than the regime of the maximum possible value of a posit ($2^{ps-2}$) or
smaller than the regime of the minimum value ($2^{2-ps}$). In these cases, $BP$
is rounded to the maximum or minimum possible value, respectively. Otherwise,
the values of $rn, rs,$ and  $nrs$, which is the number of bits that do not
represent regime bits, are computed and the regime bits are set. Using a buffer
of size $3 \cdot ps$, the fraction is shifted left with $2 \cdot ps-fs$, then
concatenated with the exponent and shifted with $ps - es$ such that the most
significant bit of the exponent is at index $3 \cdot ps$. We take the regime
bits and this concatenation of exponent and fraction shifted with $nrs$ and
store them in $BP$.

\begin{algorithm}[tp]
\caption{Posit Multiplier}
\label{alg:posit_multiplier}
\SetKwInOut{Input}{input}\SetKwInOut{Output}{output}
\Input{$P_{1}, P_{2}$}
\Output{$P_{3}$}
\BlankLine
\lIf {$(P_{1}.sn=1 \myand P_{1}.s=1) \myor (P_{2}.sn=1 \myand P_{2}.s=1)$}
{
$P_{3} \leftarrow NaR$
}
\lElseIf{$(P_{1}.sn=1 \myand P_{1}.s=0) \myor (P_{2}.sn=1 \myand P_{2}.s=0)$}
{
$P_{3} \leftarrow 0$
}
\Else
{
$P_{3}.s \leftarrow P_{1}.s \oplus P_{2}.s$\;
\tcc{\small{$\oplus$ represents exclusive or (xor)}}
$P_{3}.sn \leftarrow 0$\;
$P_{3}.k \leftarrow P_{1}.k + P_{2}.k$\;
$P_{3}.e \leftarrow P_{1}.e + P_{2}.e$\;
$P_{3}.es \leftarrow P_{1}.es$\;
$P_{3}.fs \leftarrow P_{1}.fs + P_{2}.fs$\;
$P_{3}.f  \leftarrow (P_{1}.f \cdot P_{2}.f)$\;
$P_{3}.bm = 0$\;
}
\end{algorithm}

Next, Algorithm~\ref{alg:posit_encode} rounds the posit to the nearest value
using the tie to even rule. The first bit which could not be represented in the
binary representation is stored in $b_{n+1}$. All the bits after it are $or$-ed,
and the value is stored in $bm$ if there is a bit with value 1 after $b_{n+1}$.
If both $b_{n+1}$ and $bm$ are 1 or if $bm$ is 0 and the last bit of the posit
binary representation is 1 (odd posit number), then we add 1 to the $BP$. At the
end, if $s$ is set, we take the two's complement of $BP$ as the final result.

\begin{algorithm}[tp]
\caption{Posit Divider}
\label{alg:posit_divider}
\SetKwInOut{Input}{input}\SetKwInOut{Output}{output}
\Input{$P_{1}, P_{2}$}
\Output{$P_{3}$}
\BlankLine
\lIf {$(P_{1}.sn=1 \myand P_{1}.s=1) \myor (P_{2}.sn=1 \myand P_{2}.s=1)$}
{
$P_{3} \leftarrow NaR$
}
\lElseIf{$P_{2}.sn=1 \myand P_{2}.s=0$}
{
$P_{3} \leftarrow NaR$
}
\lElseIf{$P_{1}.sn=1 \myand P_{1}.s=0$}
{
$P_{3} \leftarrow 0$
}
\Else
{
$P_{3}.s \leftarrow P_{1}.s \oplus P_{2}.s$\;
$P_{3}.sn \leftarrow 0$\;
$P_{3}.k \leftarrow P_{1}.k - P_{2}.k$\;
$P_{3}.es \leftarrow P_{1}.es$\;
\If{$P_{2}.e > P_{1}.e$}
{
$P_{3}.e \leftarrow P_{1}.e + (1 \ll P_{3}.es)- P_{2}.e$\;
$P_{3}.k \leftarrow P_{3}.k - 1$\;
}
\lElse
{
$P_{3}.e \leftarrow P_{1}.e - P_{2}.e$
}
$P_{3}.fs \leftarrow (P_{1}.fs + ps) - P_{2}.fs$\;
$P_{3}.f  \leftarrow ( (P_{1}.f \ll ps) / P_{2}.f)$\;
$P_{3}.bm \leftarrow ( (P_{1}.f \ll ps) \% P_{2}.f)$\;
}
\end{algorithm}

\begin{algorithm}
\caption{Posit SQRT}
\label{alg:posit_sqrt}
\SetKwInOut{Input}{input}\SetKwInOut{Output}{output}
\Input{$P_{1}$}
\Output{$P_{2}$}
\BlankLine

\lIf {$P_{1}.sn=1 \myand P_{1}.s=1$}
{
$P_{2} \leftarrow NaR$
}
\lElseIf{$P_{1}.sn=1 \myand P_{1}.s=0$}
{
$P_{2} \leftarrow 0$
}
\lElseIf{$P_{1}.sn=0 \myand P_{1}.s=1$}
{
$P_{2} \leftarrow NaR$
}
\Else
{
$P_{2}.s \leftarrow 0$\;
$P_{2}.sn \leftarrow 0$\;
$P_{2}.k \leftarrow P_{1}.k \gg 1$\;
$P_{2}.es \leftarrow P_{1}.es$\;
$P_{2}.f, P_{2}.bm \leftarrow $ UINT\_SQRT($P_{1}.f \gg ( (P_{1}.e \& 1) + (P_{1}.fs \& 1))$)\;
$P_{2}.e \leftarrow (P_{1}.e  + (P_{1}.e \& 1)) \gg 1$\;
$P_{2}.fs \leftarrow (P_{1}.fs - (P_{1}.fs \& 1)) \gg 1$\;

}
\end{algorithm}

\begin{algorithm}[tp]
\caption{Unsigned Integer SQRT (UINT\_SQRT)}
\label{alg:posit_uint_sqrt}
\SetKwFunction{Reverse}{Reverse}
\SetKwFunction{BitSize}{BitSize}
\SetKwInOut{Input}{input}\SetKwInOut{Output}{output}
\Input{$D$}
\Output{$Q, R$}
\BlankLine

$size \leftarrow$ \BitSize{$D$}\;
$Q \leftarrow 0$\;
$R \leftarrow 0$\;
$i \leftarrow size/2 - 1$\;
\For{$i \geq 0$}
{
$t_R \leftarrow (R \ll 2) | ((D \gg (2*i)) \& 3)$\;
\lIf{$R \geq 0$}
{$R \leftarrow t_R - ((Q \ll 2) | 1)$}
\lElse
{$R \leftarrow t_R + ((Q \ll 2) | 3)$}
\lIf{$R \geq 0$}
{$Q \leftarrow (Q \ll 1) | 1$}
\lElse
{$Q \leftarrow Q \ll 1$}
$i\leftarrow i-1$
}
\lIf{$R < 0$}
{
$R \leftarrow R + ((Q \ll 2) | 1)$
}
\end{algorithm}

\textbf{Adder, Subtractor, Multiplier and Divider.} The Posit Adder/Subtractor
presented in Algorithm~\ref{alg:posit_adder} takes two posit numbers, $P_{1}$
and $P_{2}$, the operation $op$ where 0 and 1 respectively represent add and
subtract, and computes the result, $P_{3}$. Before performing the actual
operation, an auxiliary selector, \texttt{PositAddSubSelector()}, is used to set
the actual operation, the order of the operands and the sign of the result. In
particular, we ensure that the first operand is the one with the highest
absolute value. The Adder/Subtractor checks for special cases first. If the
operands are not special posits, it computes the result and sets the extra bit,
$P_{3}.bm$.

The Posit Multiplier and Divider, presented respectively in
Algorithm~\ref{alg:posit_multiplier} and Algorithm~\ref{alg:posit_divider}, take
two posit numbers $P_{1}$ and $P_{2}$, and compute the result $P_{3}$.
Both algorithms check for special posits first, before continuing to the normal
case.

The square root extraction (SQRT) is usually done in hardware using restoring,
non-restoring~\cite{posit_sqrt_c5}, look-up table (LUT), or
approximation~\cite{posit_sqrt_c9}. We adapt a non-restoring algorithm from
\cite{posit_sqrt_c6} because it has a lower delay compared to other approaches.
We present our implementation in Algorithm~\ref{alg:posit_uint_sqrt} and the
wrapper SQRT instruction in Algorithm~\ref{alg:posit_sqrt}. The wrapper function
checks for special cases, such as values $0$, $NaR$, and negative values. For
the SQRT of a negative value, our implementation returns a $NaR$. Next, if the
input is positive, the exponent and fraction size of the result are half those
of the input. Hence, we use right-shift operations for halving and we check if
the exponent and fraction sizes are odd. If they are odd (i.e., $P_1.e\&1$ is
$1$ and $P_1.fs\&1$ is $1$), we multiply and divide, respectively, them by 2 and
we reverse the operations on the fraction. That is, we divide the fraction by 1,
2, or 4, depending on the parity of the exponent and fraction size. Lastly, we
use the non-restoring algorithm to extract the square root of the updated
fraction value, which is a positive integer.

Algorithm~\ref{alg:posit_uint_sqrt} computes the square root $Q$ and reminder
$R$ of an integer value $D$, such that $D = Q^2 + R$. The key idea is to advance
at each iteration with two digits from $D$ plus the reminder of the previous
iteration and to compute the coresponding digit in $Q$. If the reminder is
positive, the corresponding digit of $Q$ is 1, otherwise it is 0. At the end, if
the reminder is negative, we restore it to its previous positive value
by adding $(Q\ll2)|1$.

\textbf{Elasticity.} We define \textit{elasticity} as the capability of a
hardware-software system to adapt its resources to the workload. In our case,
elasticity means that \pau uses the most suitable posit size for a given
workload. Furthermore, elasticity could manifest offline, when the system is
configured or implemented, or online, during workload execution.
We leave online elasticity for future work and focus on offline elasticity in
this paper. From the hardware's perspective, offline elasticity requires that
our execution unit supports flexible posit size. Indeed, our \pau supports any
posit and exponent size. From the software's perspective, offline elasticity
requires that we identify the minimum posit size that achieves the targeted
accuracy or leads to correct results. Intuitively, identifying the range of the
real numbers used during the execution of a given workload should be a necessary
step in determining the most suitable posit size. However, we shall see in
Section~\ref{sec:eval_elastic} that this is not always the case.

\subsection{Software Support}
\label{sec:software}

On the software side, there is very limited support for the posit format. For
example, the widely-used \textit{gcc} compiler and the GNU C library do not
support posits. Given this limitation, we devise two alternatives to execute
programs with floating-point numbers on the \pau without compiler support.
The first alternative is to add a hardware conversion unit activated each time a
floating-point value is loaded from and stored to the memory. In such an
implementation, highlighted in Figure~\ref{fig:rocketchip_pipeline} under the
Memory pipeline stage, the program, memory, and caches are working with IEEE 754
representations, while the core is working with posits. This is a flexible
solution but it has two disadvantages due to frequent conversions, namely (i)
low efficiency and (ii) low accuracy. We shall see below that the loss of
accuracy is significant. The second alternative is to replace the IEEE 754 float
representation with posit representation directly in the high-level code or in
the binary. This alternative is less flexible but exhibits better efficiency. We
use this second approach in our evaluation since it yields better results.

\begin{lstlisting}[float=tp,style=customc,frame=single,caption=C code for
estimating Euler’s number ($e$) with floats or posits,label={lst:ecode}]
#ifdef POSIT_32
*((uint32_t*)&one) = posit_one;
*((uint32_t*)&two) = posit_two;
*((uint32_t*)&e) = posit_two;
*((uint32_t*)&k) = posit_two;
*((uint32_t*)&fact) = posit_one;
#else
*((uint32_t*)&one) = fp32_one;
*((uint32_t*)&two) = fp32_two;
*((uint32_t*)&e) = fp32_two;
*((uint32_t*)&k) = fp32_two;
*((uint32_t*)&fact) = fp32_one;
#endif /* POSIT_32 */
int i;
for (i = 2; i < N; i++) {
fact = fact / k;
k = k + one;
e = e + fact;
}
\end{lstlisting}

Next, we present an example to illustrate our second approach. This approach
consists of loading different binary values in the floating-point constants, as
shown in Listing~\ref{lst:ecode} for Euler's number computation using numerical
series. In this listing, all arithmetic operations are done with variables
because immediate operands are generated in \fpThirtyTwo format by the compiler,
in this case by \textit{riscv64-unknown-elf-gcc}. Hence, any immediate operand in
\fpThirtyTwo format breaks our posit implementation. Instead, we load the
variables, which are memory locations, with \fpThirtyTwo or posit, in this case
32-bit posits with 3-bit exponents. This representation solves the challenge of
not having compiler support for posits and allows us to use RISC-V extension F
ISA. Moreover, the programming approach shown in Listing~\ref{lst:ecode}
results in generating near-identical assembly code for \fpThirtyTwo and posit,
except for the values of the constants. These identical assembly footprints
ensure a fair cycle-level comparison between \fpThirtyTwo and posit.

\begin{figure}[tbp] \centering
\includegraphics[width=0.47\textwidth]{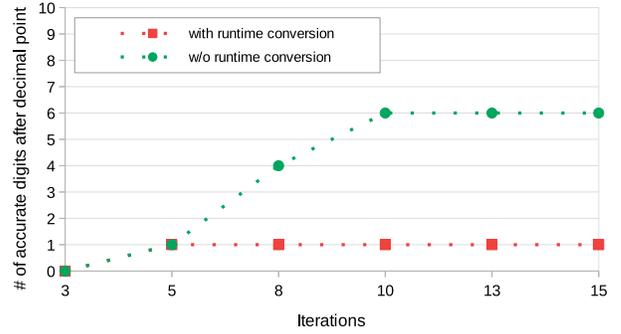}
\caption{Loss of accuracy in Euler's number computation with frequent
conversion between \fpThirtyTwo and \positThirtyTwo}
\label{fig:euler_accuracy_decenc}
\end{figure}

\textbf{Impact of Runtime Conversion.} Lastly, we present empirical evidence
that the first alternative of converting between \fpThirtyTwo and posit at
runtime, in the hardware, leads to loss of accuracy. Taking Euler's number
computation as an example, shown in Listing~\ref{lst:ecode}, we emulate runtime
conversion by encoding 32-bit IEEE 754 single-precision floating-points
(\fpThirtyTwo) to 32-bit posits with 3-bit exponents (\positThirtyTwo) before
each iteration, and decoding \positThirtyTwo to \fpThirtyTwo after each
iteration. This represents the runtime behavior of having \fpThirtyTwo values in
the memory, including cache, and \positThirtyTwo values in the CPU's core
registers. As shown in Figure~\ref{fig:euler_accuracy_decenc}, the loss of
accuracy is drastic. With runtime conversions, only one digit of the fraction
is accurate, i.e. $e = 2.7$, while directly loading into the memory and
operating with \positThirtyTwo leads to six accurate fraction digits, which is
the same as the accuracy achieved by \fpThirtyTwo. In the remainder of this
paper, we are using the approach highlighted in Listing~\ref{lst:ecode} since it
leads to higher accuracy.

\section{Evaluation}
\label{sec:eval}

In this section we evaluate and analyze our approach based on accuracy,
efficiency, estimated area and power.

\subsection{Setup}
\label{sec:setup}

We compare the original 32-bit FPU of Rocket Chip which claims to implement the
IEEE 754 standard (\textbf{\fpThirtyTwo}) with our \pau operating with posits of
three bit widths, namely 8-bit with 1-bit exponent denoted by
\textbf{\positEight} or \textbf{P8}, 16-bit with 2-bit exponent denoted by
\textbf{\positSixteen} or \textbf{P16}, and 32-bit with 3-bit exponent denoted
by \textbf{\positThirtyTwo} or \textbf{P32}. We wrote the high-level code for
\pau in Chisel, integrated it with Rocket Chip~\cite{asanovic2016rocket},
and used SiFive's Freedom E310\footnote{https://github.com/sifive/freedom}
development platform to implement and synthesize our code to run on an Arty
A7-100T FPGA.
The original Rocket Chip with FPU and Rocket Chip with POSAR run at the same frequency.


Even if the Arty A7-100T FPGA represents the high-end of its family, it still
exhibits a serious limitation in terms of available data memory which hinders
the execution of full-fledged ML models. The 4,860 Kb (kilobits) block RAM of
Arty A7-100T limits the data memory size of our Rocket Chip core to 512 kB
(kilobytes). However, typical ML models have thousands or millions of weights
which take several MB of memory. Even for the classic matrix multiplication, we
can only run it on square matrices of size up to 182. In our future work, we
plan to address this limitation by linking Rocket Chip to the 256 MB of RAM
available on the Arty A7-100T. However, in this paper, we can only use a maximum
of 512 kB of main memory.

\subsection{Benchmarks}
\label{sec:bench}

To evaluate our approach, we select benchmarks that use floating-point
operations. We organize these benchmarks into three levels as follows. Level one
benchmarks are used to evaluate both the accuracy and efficiency, in terms of
cycles, of our \pau versus the original IEEE 754 FPU of Rocket Chip. These
benchmarks represent the computation of well-known mathematical constants using
series and sequences. In particular, we compute the constants $\pi$ and $e$
(Euler's number), using numerical series, as shown in
Table~\ref{tab:level_one_accuracy}. For $\pi$, we use Leibniz and Nilakantha
series~\cite{pi_formula}. Since Leibniz series converges slowly, we run it for
two million iterations. In contrast, Nilakantha series converges faster, thus,
we run it for 200 iterations. For $e$, we use Euler's series which is
fast-converging, thus, we run it for 20 iterations. In addition to $\pi$ and
$e$, we also compute $sin(1)$ with 10 iterations.

Level two consists of kernels that are typically used in ML
applications~\cite{pudiannao15}, as summarized in Table~\ref{tab:level_two}.
For these kernels, we evaluate the efficiency of our \pau versus the FPU in
terms of cycles. The correctness of the results is checked against reference
outputs. Next, we briefly describe each kernel. Matrix Multiplication
(\textbf{MM}) implements the multiplication of two square matrices which is
often used in ML and HPC workloads. In our testbed, we can accommodate matrices
of size up to $n = 182$. k-means (\textbf{KM}) groups a set of multi-dimensional
points into $k$ groups, or clusters, based on their Euclidean distance. KM is
often used in ML and data analytics applications. k-nearest neighbors
(\textbf{KNN}) classifies a multi-dimensional point based on the Euclidean
distance to its $k$ nearest neighbors. Linear Regression (\textbf{LR}) is a
kernel used in ML and data analytics. We implement Multivariate Linear
Regression which consists of matrix and vector operations. Naive Bayes
(\textbf{NB}) implements a simple Bayesian model. The Classification (or
Decision) Tree (\textbf{CT}) kernel is used in ML and data analytics to
represent a target variable based on some input attributes. We implement both
the creation (training) and usage (inference) of CT. We use Iris
dataset\footnote{\url{https://archive.ics.uci.edu/ml/datasets/iris}} as input
for level two benchmarks, except MM. This dataset consists of $n = 150$ data
points with $m = 4$ dimensions representing flowers. These points belong to $k =
3$ classes.

\begin{figure}[tp]
\centering
\includegraphics[width=0.33\textwidth]{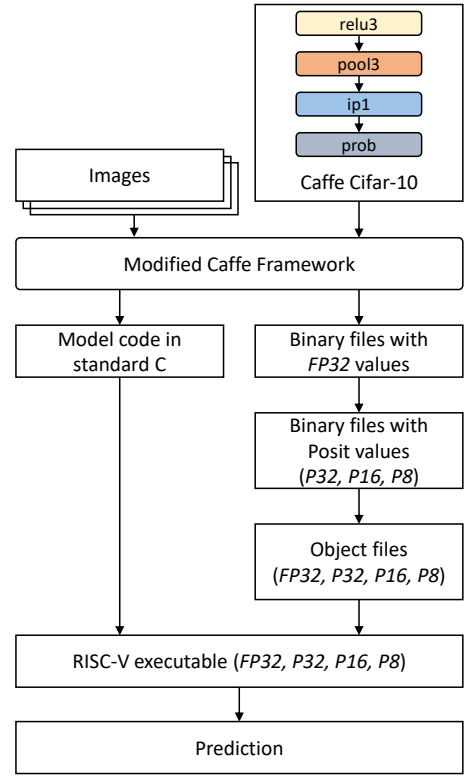}
\caption{CNN evaluation flow}
\label{fig:cnn_riscv}
\end{figure}

Level three is represented by one NAS Parallel Benchmark (NPB)~\cite{NPB_2011}
scientific application and one Convolutional Neural Network (CNN) ML inference
application. Specifically, we selected Block Tri-diagonal (\textbf{BT}) solver
from NPB and we converted all floating-point variables to 32-bit float. We use
the verification threshold error, \textit{epsilon} ($\epsilon$), as a measure of
accuracy. That is, a smaller $\epsilon$ corresponds to a higher accuracy.
Next, we use a Convolutional Neural Network (CNN) implemented in
Caffe~\cite{caffe_14} and trained on
Cifar-10\footnote{\url{https://www.cs.toronto.edu/~kriz/cifar.html}} dataset, as
depicted in Figure~\ref{fig:cnn_riscv}. This CNN has 14 layers and the
parameters file has a size of 351 kB. However, we cannot accommodate this model
on our testbed with limited memory size. Hence, we take only the last four
layers of this CNN, starting from \textit{relu3} as shown in
Figure~\ref{fig:cnn_riscv}, and generate standard C code with static memory
allocations in order to run it on the bare-metal SiFive's Freedom E310.
By instrumenting the Caffe framework, we collect all the parameters and the
input of \textit{relu3} layer as binary files with \fpThirtyTwo values.
We then convert these binaries to all three posit sizes, namely P8, P16, and
P32, transform them into objects and link them with the generated C code to get
the final RISC-V executable, as shown in Figure~\ref{fig:cnn_riscv}. We then run
the validation on all 10,000 images of Cifar-10 test dataset by running the
executables on the Arty A7-100T FPGA. The prediction results are compared
against the reference execution on an x86/64 host.


\subsection{Accuracy and Efficiency}
\label{sec:accu}

\textbf{Level One.} We evaluate the accuracy and efficiency of posit in
comparison with 32-bit, single-precision IEEE 754 floating-point (\fpThirtyTwo),
using level one benchmarks summarized in Table~\ref{tab:level_one_accuracy} and
Table~\ref{tab:level_one_efficiency}. The accuracy is measured in terms of exact
fraction digits compared to the reference value of the mathematical constant.
The efficiency represents the number of cycles taken by Rocket Chip running
on the FPGA to execute the meaningful section of the program. For posits,
we compute the speedup with respect to the \fpThirtyTwo execution. We note that
we use 64-bit, double-precision IEEE 754 floating-point in our evaluation scripts.
This is because any posit can be accurately represented by an IEEE 754 float of bigger
size~\cite{de2019posits}.

The results presented in Table~\ref{tab:level_one_accuracy} show that
\positThirtyTwo achieves similar or better accuracy compared to \fpThirtyTwo.
Moreover, \positThirtyTwo achieves a speedup of 1.3 compared to \fpThirtyTwo FPU,
when $\pi$ with Leibniz series is computed, as shown in Table~\ref{tab:level_one_efficiency}.
The accuracy of small posit representations, such as \positEight, is low when
estimating numerical series. This is expected if we consider the internals of
posit representation. Taking $e = 2.7182\ldots$ as example, we first observe
that the closest \positEight numbers are 2.625 (\texttt{0x55}) and 2.75
(\texttt{0x56}). That is, one cannot get better accuracy for $e$ than these
two values. Second, Euler series leads to an issue regarding the storage in
\positEight of the factorial which grows very fast. The maximum value that a
\positEight can represent is 192, which is less than $6!$. Hence, the accuracy
of Euler's series becomes worst when the number of iterations grows. For example,
when $N = 4$, we get $e = 2.75$, but when $N = 6$ we get $e = 3$.

\begin{figure}[tbp] \centering
\includegraphics[width=0.48\textwidth]{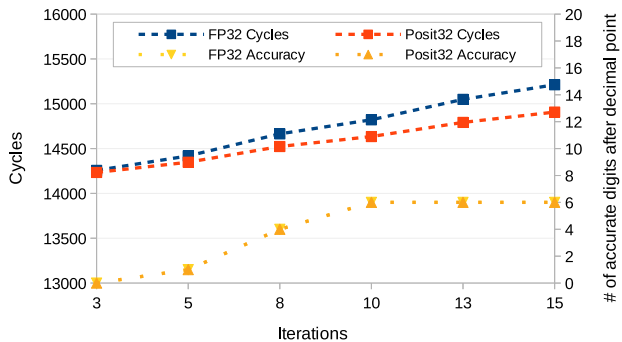}
\caption{Accuracy and efficiency of Euler's number computation using
\fpThirtyTwo and \positThirtyTwo}
\label{fig:euler_accuracy_cycles}
\end{figure}

\begin{table*}[tp]
\centering
\caption{Accuracy (Level One Benchmarks)}
\label{tab:level_one_accuracy}
\resizebox{0.68\textwidth}{!} {
\begin{tabular}{|l|r|r|r|r|r|r|r|r|r|}
\hline
\multirow{3}{*}{\textbf{Application}} & \multirow{3}{*}{\textbf{Iterations}}
& \multicolumn{8}{c|}{\textbf{Accuracy}} \\
& & \multicolumn{8}{c|}{[actual value $|$ number of exact fraction digits]} \\
\cline{3-10} & & \multicolumn{2}{c|}{\textbf{\fpThirtyTwo}} &
\multicolumn{2}{c|}{\textbf{\positEight}} &
\multicolumn{2}{c|}{\textbf{\positSixteen}} &
\multicolumn{2}{c|}{\textbf{\positThirtyTwo}} \\
\hline
\hline
$\pi$ (Leibniz) & 2,000,000 & 3.14159 & 5 & 3.5 & 0 & 3.14 & 2 & 3.14159 & 5 \\
$\pi$ (Nilakantha) & 200 & 3.1415929 & 6 & 3.125 & 1 & 3.141 & 3 & 3.1415922 & 6
\\
$e$ (Euler) & 20 & 2.7182819 & 6 & 2.625 & 0 & 2.718 & 3 & 2.7182817 & 6 \\
$sin(1)$ & 10 & 0.8414709 & 7 & 0.78 & 0 & 0.8413 & 3 & 0.84147098 & 8 \\
\hline
\end{tabular}
}
\tablevspace
\end{table*}

\begin{table*}[tp]
\centering
\caption{Efficiency (Level One Benchmarks)}
\label{tab:level_one_efficiency}
\resizebox{0.82\textwidth}{!} {
\begin{tabular}{|l|r|r|r|r|r|r|r|r|}
\hline
\multirow{3}{*}{\textbf{Application}} & \multirow{3}{*}{\textbf{Iterations}}
& \multicolumn{7}{c|}{\textbf{Efficiency}} \\
& & \multicolumn{7}{c|}{[cycles $|$ speedup]} \\
\cline{3-9} & & \textbf{\fpThirtyTwo} &
\multicolumn{2}{c|}{\textbf{\positEight}} &
\multicolumn{2}{c|}{\textbf{\positSixteen}} &
\multicolumn{2}{c|}{\textbf{\positThirtyTwo}} \\
\hline
\hline
$\pi$ (Leibniz) & 2,000,000 & 216,022,827 & 166,022,835 & 1.30 & 166,022,829 &
1.30 & 166,022,830 & 1.30 \\
$\pi$ (Nilakantha) & 200 & 57,940 & 52,937 & 1.09 & 52,952 & 1.09 & 52,937 &
1.09
\\
$e$ (Euler) & 20 & 15,598 & 15,177 & 1.03 & 15,177 & 1.03 & 15,177 & 1.03 \\
$sin(1)$ & 10 & 16,663 & 16,270 & 1.02 & 16,273 & 1.02 & 16,298 & 1.02 \\
\hline
\end{tabular}
}
\tablevspace
\end{table*}

\begin{table*}[tp]
\centering
\caption{Efficiency (Level Two Benchmarks). Gray background means that the
result is different from the reference.}
\label{tab:level_two}
\resizebox{0.92\textwidth}{!} {
\begin{tabular}{|l|c|r|r|r|r|r|r|r|}
\hline
\multicolumn{1}{|c|}{\multirow{2}{*}{\textbf{Benchmark}}} &
\multirow{2}{*}{\textbf{Input Size}} & \multicolumn{7}{c|}{\textbf{Efficiency}
[cycles $|$ speedup]} \\
\cline{3-9} & & \textbf{\fpThirtyTwo} &
\multicolumn{2}{c|}{\textbf{\positEight}} &
\multicolumn{2}{c|}{\textbf{\positSixteen}} &
\multicolumn{2}{c|}{\textbf{\positThirtyTwo}} \\
\hline
\hline
Matrix Multiplication (MM) & n = 182 & 418,177,415 & 418,063,614 & 1.0 &
418,063,629 & 1.0 & 418,177,423 & 1.0 \\
\hline
k-means (KM) & \multirowcell{5}{Iris dataset \\ n = 150 \\ m = 4 \\ k = 3 } &
19,150,075 & \cellcolor{mGrayTbl} 18,879,618 & \cellcolor{mGrayTbl} 1.01 & 18,971,747 & 1.01 & 19,011,507 & 1.01 \\
k Nearest Neighbours (KNN) & & 151,402 & \cellcolor{mGrayTbl} 138,140 &
\cellcolor{mGrayTbl} 1.10 & 143,313 & 1.06 & 144,136 & 1.05 \\
Linear Regression (LR) & & 1,419,794 & \cellcolor{mGrayTbl} - & \cellcolor{mGrayTbl} - & \cellcolor{mGrayTbl} 1,398,782 & \cellcolor{mGrayTbl} 1.02 & 1,398,643 & 1.02 \\
Naive Bayes (NB) & & 398,254 & \cellcolor{mGrayTbl} 407,330 &
\cellcolor{mGrayTbl} 0.98 & 397,869 & 1.0 & 399,893 & 1.0 \\
Classification Tree (CT) & & 633,560 & 101,940 & 6.2 & 615,792 & 1.03 & 629,936
& 1.01 \\
\hline
\end{tabular}
}
\tablevspace
\end{table*}

Posit operations take fewer cycles to complete, thus, applications with higher
numbers of iterations exhibit better efficiency. For example, \positThirtyTwo is
30\%, 9\%, and 3\% faster than \fpThirtyTwo for $\pi$ Leibniz with two million
iterations, $\pi$ Nilakantha with 200 iterations, and $e$ with 20 iterations,
respectively. Our analysis revealed that this speedup is the result of faster
multiplication and division operations on posits. This, in turn, is the result
of simpler exception and corner case handling in posits. Intuitively, the gap in
efficiency grows with the number of iterations.
Figure~\ref{fig:euler_accuracy_cycles} shows that \positThirtyTwo achieves the
same accuracy as \fpThirtyTwo with fewer cycles as the number of iterations
increases.

\textbf{Level Two.} We observe that \positThirtyTwo and \positSixteen lead to
the same final results as \fpThirtyTwo when running level two benchmarks while
saving up to 6\% of the cycles, as shown in Table~\ref{tab:level_two}. However,
LR with \positEight and \positSixteen exhibits wrong results. In turn, the final
results are affected by the wrong value of one of the determinants computed by
the program. In fact, all the programs operating with \positEight produce wrong
results, except CT. This shows that small size posits are not suitable for some
classic ML kernels that need high numerical accuracy. This observation is in
contrast to some of the related works~\cite{jeff_fb, deep_positron_arxiv,
deep_nn_gustafson}. However, we note that our evaluation is done on a different
dataset, namely the Iris dataset. We shall see below that \positEight performs
better on a partial CNN. On the other hand, \positSixteen offers a good
alternative to 32-bit floating-point representations.

\textbf{Level Three.} For the NPB application, \positThirtyTwo achieves one level of
magnitude higher accuracy than \fpThirtyTwo. For example, setting $\epsilon =
10^{-4}$ in BT leads to successful validation when \positThirtyTwo is used. On the
other hand, \fpThirtyTwo needs $\epsilon = 10^{-3}$ in BT to pass the
validation. Moreover, \positThirtyTwo exhibits a marginal speedup compared to
\fpThirtyTwo. These results are in correlation to those of level one benchmarks.
Since there are more and diverse floating-point operations in BT compared to
level one benchmarks, the accuracy gain of \positThirtyTwo is more visible. On the
other hand, the speedup gain is not spectacular because the fraction of
operations where posit is faster than IEEE 754 is smaller. For the same reason
of very large number of operations in BT, \positEight and \positSixteen do not exhibit
good accuracy. In fact, \positEight cannot even represent accurately all the
validation reference values due to its limitted range. For example, the
validation reference value $7.38e-5$ of BT cannot be represented by \positEight
because its range stops at $2.44e-4$ (\texttt{0x1}).

When compared to the reference execution on an x86/64 host,
the Cifar-10 CNN in Figure~\ref{fig:cnn_riscv} with \fpThirtyTwo,
\positThirtyTwo and \positSixteen running on our FPGA with a Rocket Chip core
exhibit the same Top-1 accuracy as the reference model, namely 68.15\%. Even
\positEight achieves a reasonable accuracy of 62.68\%. In terms of speed, all three
posit representations are around 18\% faster compared to the execution with
\fpThirtyTwo. The results with \positSixteen and \positEight are very promising
and open-up a series of future optimizations. For example, these formats save
respectively half and three-quarters of the memory for representing inputs and
parameters compared to 32-bit \fpThirtyTwo or \positThirtyTwo. Next, by packing
two \positSixteen and four \positEight operands per instruction, we can reduce
the execution time by two and four times, respectively.

We observe that one reason why \positEight exhibits accuracy loss is due to the
out-of-range representation of some parameters or input image pixels. There is
at least one out-of-range representation for each of the 10,000 input images of
the Cifar-10 test dataset. This is in contrast to \positThirtyTwo and
\positSixteen which can represent these parameters without loss of accuracy. For
example, the minimum positive value of the weights of \textit{ip1} layer is
$0.000001119$ which cannot be represented by \positEight. The closest posit size
that can represent this value relatively accurate is \textit{Posit(15,2)} with
$0.0000011176$ (\texttt{0x10b}). We note that scaling cannot be applied for
\positEight because of the wide parameter distribution interval, that is, the
minimum positive value is $0.000001119$ and the maximum one is $87.84$.

Another source of accuracy loss is due to underflow or overflow at runtime. For
example, \textit{prob} layer includes exponentiation, among other operations. On
\positEight, exponentiation can easily result in underflow or overflow. To test
this hypothesis, we keep the parameters in 8-bit posit format in memory but we
employ the \pau with \positSixteen and convert between these two formats at
runtime. The result is better than expected because the Top-1 accuracy of this
approach is 68.47\%, a bit higher than the accuracy of the reference execution
on \fpThirtyTwo. This result confirms our hypothesis that the main source of
inaccuracy of \positEight is at runtime and shows that using a hybrid approach
with posits of different size can save memory without losing accuracy.

\subsection{Dynamic Range and Elasticity}
\label{sec:eval_elastic}

\begin{table}[tp]
\caption{Dynamic floating-point range of all benchmarks}
\label{table:ranges}
\centering
\begin{tabular}{|l|r|r|}
\hline
\multicolumn{1}{|c}{\multirow{3}{*}{\textbf{Benchmark}}} & \multicolumn{1}{|c|}{\textbf{Minimum}} & \multicolumn{1}{c|}{\textbf{Maximum}} \\
& \multicolumn{1}{c|}{\textbf{Value}} & \multicolumn{1}{c|}{\textbf{Value}} \\
& \multicolumn{1}{c|}{in $(0,1]$} & \multicolumn{1}{c|}{in $[1,\infty)$} \\
\hline
\hline
$\pi$ (Leibniz) & 1.0e-06 & 3,999,999 \\
$\pi$ (Nilakantha) & 6.2e-08 & 64,480,800 \\
$e$ (Euler) & 8.22e-18 & 20 \\
$sin(1)$ & 1.96e-20 & 9.223e+18 \\
\hline
KM & 2.22e-16 & 245.8 \\
KNN & 9.99e-03 & 395,090 \\
LR & 0.01 & 140,690,992 \\
NB & 1.49e-06 & 150 \\
CT & 2.53e-14 & 4 \\
\hline
CNN & 1.4E-45 & 3,184,598,272 \\
\hline
\end{tabular}
\end{table}

To help developers analyze the floating-point values used during the execution
of a program, we implemented an instrumentation tool based on
DynamoRIO~\cite{dynamorio_cgo03}. This tool takes a binary and inspects the
registers and memory locations involved in \fpThirtyTwo instructions. This
instrumentation does not require high-level code changes and can be done on an
x86/64 host. At the moment, we cannot perform this analysis directly on the
RISC-V ISA since DynamoRIO does not support it yet.

Using this tool, we determine the absolute minimum value in the interval $(0,1]$
and the absolute maximum value in the interval $[1,\infty)$. These values
determine the minimum posit size needed to represent the real numbers involved
in the binary's execution. The results for all our benchmarks running on the
inputs stated in Table~\ref{tab:level_one_efficiency} and
Table~\ref{tab:level_two} are summarized in Table~\ref{table:ranges}. We note
that the minimum values higher than zero that can be represented by \positEight,
\positSixteen, and \positThirtyTwo are $2^{-10}$, $2^{-48}$, and $2^{-216}$,
respectively. The maximum values that can be represented by the same posit sizes
are $2^{9}$, $2^{47}$, and $2^{215}$, respectively.

Interestingly, we do not observe a strong correlation between the dynamic range
and wrong results. For example, \positSixteen covers the dynamic range of almost
all benchmarks, except $e$, $sin(1)$, KM, and CNN. But it produces correct
results for KM and CNN, while for $e$ and $sin(1)$ it leads to the same number
of accurate digits as for $\pi$. On the other hand, the dynamic values of LR are
in the range of \positSixteen, but the final result is wrong. Hence, it is not
sufficient to perform this analysis in order to determine the most suitable
posit size for an application. For the offline elasticity of \pau, developers
must simulate or run the application with different posit sizes and select the
most suitable size for the application in the hardware.

\subsection{Resource Utilization}
\label{sec:area}

As a proxy to the chip area taken by our implementation, we evaluate the FPGA
resource utilization of our \pau compared to the original FPU of Rocket Chip. We
evaluate the FPGA resource utilization of the entire system, namely SiFive
Freedom E310 with a Rocket Chip core that has an FPU/\pau, running on the Arty
A7-100T FPGA. While the results here denote savings in terms of resources from
an FPGA perspective, similar or even higher savings in terms of the area will be
obtained when the design is implemented on an
ASIC~\cite{ehliar2009asic,fpga-asic-gap}. Savings in area directly relate to
savings in both static and dynamic power and thus are important for low-power
constrained applications such as IoT-based edge devices.
Table~\ref{table:area_fpga} shows the utilization of the different FPGA
resources with respect to both posit and \fpThirtyTwo implementations.

We observe that all the implementations use the same amount of memory resources
(Shift-register Look up table -- SRL, LUTRAM, and BRAM) which indicates that the
comparison involves only the modified FPU with the rest of the system being the
same across all implementations. For significant savings in area and power
without much loss in accuracy \positSixteen seems to be a viable option that
saves almost 50\% of the DSPs which translate to the multiply-accumulate (MAC)
units in an ASIC flow. These savings in area should translate to a 50\% drop in
dynamic power as the MACs account for a higher power compared to flops or other
logic~\cite{garland2018low}. In contrast, \positThirtyTwo uses 30\% more LUTs
and 27\% more DSPs compared to \fpThirtyTwo. These results are worse than those
reported in~\cite{chaurasiya2018parameterized}, which needs only 4\% more LUTs
compared to the FPU, but similar to the ones reported in~\cite{8342187}. On the
other hand, we note that the original FPU of Rocket Chip is a work-in-progress.
That is, it may not implement all the corner cases of IEEE 754 standard.
Nonetheless, the higher resource utilization of \positThirtyTwo may be
counterbalanced by its speedup which can lead to higher time and energy
efficiency compared to \fpThirtyTwo.

\begin{table}[tp] \centering
\caption{FPGA Resource Utilization of the Entire Rocket Chip on SiFive Freedom
E310}
\label{table:area_fpga}
\resizebox{0.485\textwidth}{!} {
\begin{tabular}{|l|r|r|r|r|}
\hline
\textbf{Resource} & \textbf{\fpThirtyTwo} & \textbf{\positEight}
&\textbf{\positSixteen} &\textbf{\positThirtyTwo} \\
\hline
\hline
\textbf{Logic LUT} & 29,335 & 19,367 (-34\%) & 25,598 (-13\%) & 38,155 (+30\%)
\\
\textbf{FF} & 14,756 & 11,596 (-21\%) & 12,031 (-19\%) & 12,951 (-12\%) \\
\textbf{DSP} & 15 & 5 (-67\%) & 8 (-47\%) & 19 (+27\%) \\
\textbf{SRL} & 58 & 60 & 60 & 60 \\
\textbf{LUTRAM} & 924 & 924 & 924 & 924\\
\textbf{BRAM} & 14 & 14 & 14 & 14\\
\hline
\end{tabular}
}
\vspace{10pt}
\end{table}

\subsection{Power and Energy}
\label{sec:power}

We measure the average power of the FPGA as a proxy to the energy efficiency of
our \pau versus the original FPU of Rocket Chip. For these measurements, we
connect a Yokogawa power meter to the 12 V DC input of the FPGA. This power
meter allows reading the power once per second. We run only $\pi$ computation
using Leibniz series with two million iterations and MM with input size 182
since they exhibit sufficiently high execution time to allow us to measure the
power. The average power of the FPGA with the original IEEE 754 FPU when running
$\pi$ and MM is 1.39 W and 1.48 W, respectively. The higher power of MM is due
to the extended data memory size since MM with input size 182 cannot run on the
default 16 kB memory. When running $\pi$ on \pau with posit size 8, 16, and 32
bits, the average power is 1.38 W, 1.4 W, and 1.48 W, respectively. When running
MM, the respective average power is 1.47 W, 1.51 W, and 1.52 W. The results are
encouraging. \pau with \positThirtyTwo uses 6\% more power than the FPU, but is
30\% faster when computing $\pi$. Hence, the energy efficiency of \pau is
superior. On the other hand, \pau with \positThirtyTwo uses only 2\% more power
when running MM, while \pau with \positEight uses 1\% less power compared to the
FPU. While the power savings of posit on the FPGA are not spectacular, an ASIC
implementation should yield even higher power
savings~\cite{ehliar2009asic,fpga-asic-gap}.

\vspace{16pt}
\section{Conclusions}
\label{sec:concl}

In this paper, we explore the opportunity of replacing the traditional IEEE 754
floating-point standard with the newly-proposed posit
format~\cite{gustafson2017beating} in the context of machine learning at the
edge. We present our implementation of an \pau to replace the original FPU in a
RISC-V core. We are the first to do a thorough evaluation of posit versus
\fpThirtyTwo to determine (i) whether hardware or software conversion between
posit and \fpThirtyTwo is better, (ii) the time-energy performance for both
mathematical and ML kernels, and (iii) insights on choosing the bit-width of
posits for different types of applications. We evaluate our implementation on an
FPGA using the SiFive Freedom E310 platform. We compare the accuracy, efficiency
in terms of cycles, FPGA resource utilization and power of three posit sizes
compared to 32-bit, single-precision IEEE 754 floats.

We find that 8-bit posits are not producing the required accuracy to replace
\fpThirtyTwo in classic ML applications. On the other hand, 32-bit posit are not
exhibiting spectacular improvements in terms of efficiency over \fpThirtyTwo.
While they achieve the same or higher accuracy and can speedup the execution
when the program has multiplications and divisions, they need around 30\% more
FPGA resources and use 6\% more power compared to \fpThirtyTwo. However, 16-bit
posits exhibit the best results. Even if they exhibit lower accuracy in
scientific computations, they produce correct results for most of ML kernels and
applications, while requiring less area and power compared to \fpThirtyTwo. For
example, \positSixteen achieves the same Top-1 accuracy as \fpThirtyTwo on a
Cifar-10 CNN while exhibiting 18\% speedup.

\balance

\vspace{16pt}
\bibliography{main}




\end{document}